\documentstyle[aps,pra,graphicx]{revtex}

\begin{document}

\title{Coherent control using adaptive learning algorithms}
\author{B. J. Pearson, J. L. White, T. C. Weinacht, and P. H. Bucksbaum}
\address{Physics Department, University of Michigan, Ann Arbor, MI 48109-1120}
\date{\today}
\maketitle

\begin{abstract}

We have constructed an automated learning apparatus to control
quantum systems. By directing intense shaped ultrafast laser
pulses into a variety of samples and using a measurement of the
system as a feedback signal, we are able to reshape the laser
pulses to direct the system into a desired state. The feedback
signal is the input to an adaptive learning algorithm. This
algorithm programs a computer-controlled, acousto-optic modulator
pulse shaper. The learning algorithm generates new shaped laser
pulses based on the success of previous pulses in achieving a
predetermined goal.

\end{abstract}


\section{\noindent Introduction}

An original impetus for coherent control was mode-selective laser
photochemistry. This dream of exciting a specific bond with a
laser has dimmed for various reasons. Control is hampered by fast
relaxation, complicated mode structure, imperfect knowledge of the
inter-nuclear potentials along arbitrary coordinates, and the
distorting influence of the strong external light fields. Recent
theoretical \cite{rabitz} and experimental advances have led to
successful control of molecular ionization and dissociation
\cite{gordon,gerber-science}, atomic and molecular fluorescence
\cite{silberberg-nature,wilson,zewail}, and excitations or
electrical currents in semiconductors
\cite{heberle,bonadeo,vandriel}. Control has been predicted and
demonstrated in several systems that have only two interfering
pathways \cite{gordon,bonadeo,vandriel}.

Strongly coupled systems such as large molecules in condensed
phase are so complicated that it is nearly impossible to calculate
optimal pulse shapes in advance. Recent efforts have used
experimental feedback as suggested by Judson and Rabitz
\cite{rabitz} to determine the optimal optical pulse shape to
achieve a particular control goal \cite{gerber-science,wilson}. In
the absence of detailed knowledge of the system Hamiltonian, an
algorithm must be used for selecting new pulse shapes inside the
feedback loop.

This paper presents a detailed investigation of a learning
strategy incorporating a modified genetic algorithm (GA)\cite{GA}
in order to discover control pathways in complicated physical
systems. Experiments have been carried out in a variety of
molecular systems whose common feature is the existence of
observable physical properties that depended on the incident
(shaped) light. Several systems are described. In the gas phase
we explore nonlinear control of ionization channels in diatomic
sodium. Several liquid phase experiments are also described,
including control of self-phase modulation in $CCl_{4}$ and
excitation of vibrational modes in methanol and benzene.

In addition to traditional GA search strategies, we incorporate
other search methods that all run concurrently. The algorithm
adapts itself during a run in order to search the phase space most
efficiently. Because of this, it is possible to obtain physical
insight into the problem under investigation not only through the
pulse-shape solutions, but also by how the algorithm arrived at
those solutions. We demonstrate that our adaptive evolutionary
algorithm is capable of controlling a diverse array of systems.

\section{\noindent Experimental Setup}

\begin{figure}[tb]
\centering
\includegraphics [width=8.6cm, height=8.6cm, angle=0]{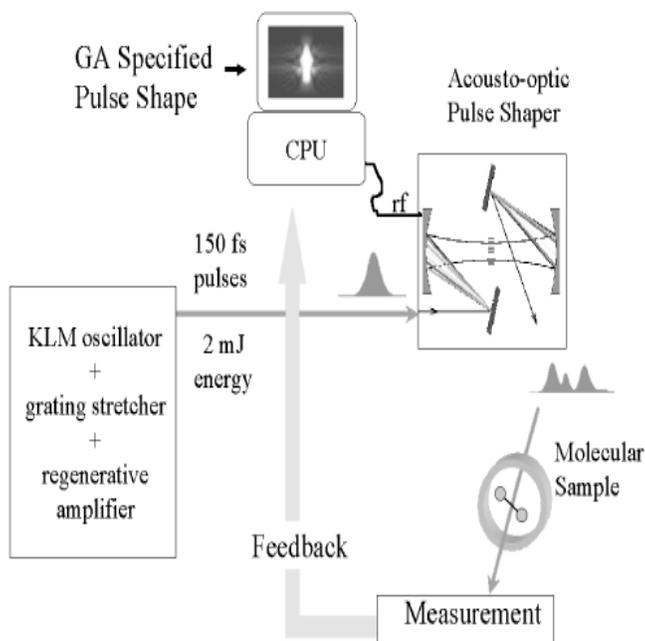}
\caption{Diagram of the experimental setup.  The shaped laser
pulses interact with a molecular sample, and a single-valued
feedback function is returned to the computer.  The algorithm uses
these values to determine the mating procedure. } \label{setup}
\end{figure}

The control light field is made with a Kerr lens mode-locked
(KLM) titanium sapphire laser that produces $100$ fs pulses with
about $5$ nJ of energy at a central wavelength of $790$ nm (see
Fig. \ref{setup}). The pulses are temporally dispersed to $150$ ps
in a single grating expander and amplified to $2$ mJ in a
regenerative amplifier at $10$ Hz. The output of the amplifier is
split to form two beams. One beam is sent to the pulse shaper,
and the other is used as an unshaped reference for spectral
interferometry measurements of the shaped pulses \cite{specint}.
The pulse shaper consists of a zero dispersion stretcher with an
acousto-optic modulator (AOM) in the Fourier plane, where
different colors in the light pulse map to different positions on
the modulator \cite{warren}. The AOM carries a shaped acoustic
traveling wave, which diffracts different colors with different
phases and amplitudes. These colors are reassembled at the output
of the pulse shaper, yielding a temporally shaped laser pulse. In
order to compensate for the low efficiency of the pulse shaper
($10-15\%$), we reamplify the shaped pulses in a low-gain
multipass amplifier. The pulses are then compressed in a single
grating compressor. The resulting shaped laser pulses are then
directed into the molecular sample to be studied, and a
predetermined feedback signal is monitored.

\section{The Learning Algorithm}

\subsection{Constructing the Algorithm}

In previous work we demonstrated control of simple quantum systems
using feedback \cite{nature}. Now we wish to control more complex
systems where the optimal pulse shape is not known {\it a
priori}. The pulse shaper has 8 bit control over the amplitude and
phase of 100 different frequency components, so there are
$2^{1600}$ different pulse shapes. The physical system couples
different frequencies, so they cannot each be optimized
independently. This presents a vast phase space for the search
algorithm. The search must also be robust in the face of
experimental noise, and capable of escaping local maxima in a
rough potential energy landscape.

These requirements led us to consider a genetic algorithm, which
fulfills all of these conditions: It does not rely on any local
information about the search space, such as derivatives, and it
is capable of handling a multidimensional phase space in the
presence of noise. Previous work with adaptive feedback found
genetic algorithms (or similar evolutionary strategies
\cite{schwefel}) adept at solving similar problems
\cite{rabitz,gerber-science,motzkus}.  A GA provides a general,
robust approach to optimization, requiring little prior
information about the problem to be solved. Furthermore, the GA
can generate new pulse shapes that differ radically from the
previous pulse shapes, to quickly sample different sections of
the phase space. Our GA implementation is derived from the {\it
Handbook of Genetic Algorithms}, a standard treatise on the
subject \cite{GA}.

Genetic algorithms are a class of search and optimization
algorithms inspired by biological evolution. In evolution,
natural selection links a string of genes to the structures that
it represents. The gene string carries all of the information,
while a structure's success determines its chance of reproducing.
Nature knows nothing about the problem to be solved; gene strings
encoding successful structures merely reproduce more often.
During reproduction, the algorithm has no regard for the structure
represented by the gene string. Offspring have new gene strings
representing new, and possibly more successful, structures. By
incorporating these ideas, a GA can solve surprisingly difficult
problems. Each possible solution (structure) in the search space
is given a representation (gene string). This representation is
called an individual, and a group of individuals is called a
population.

\begin{figure}[tb]
\centering
\includegraphics [width=8.6cm, height=8.6cm, angle=0]{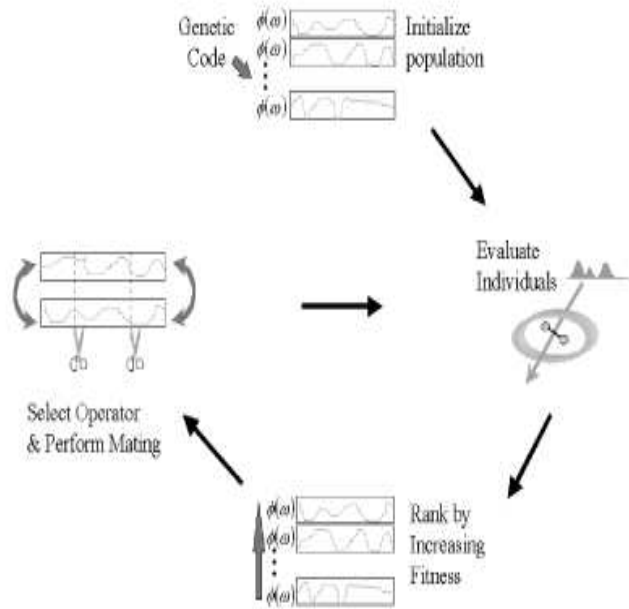}
\caption{The adaptive algorithm begins by initializing a random
population of pulse shapes.  These pulse shapes interact with the
molecular system and are ranked according to fitness. These
fitnesses, as well as the fitnesses of the various operators, are
used to select the individuals and operators to be used in the
mating.  After the new generation of pulse shapes is created, the
cycle repeats.} \label{ga-diagram}
\end{figure}

A basic example of a GA's implementation is shown in Fig.
\ref{ga-diagram}. Each individual in the population is evaluated
on a given molecular sample. The individuals are ranked from most
to least fit based on a predetermined single-valued feedback
function (the fitness). These individuals (parents) then
reproduce according to some protocol, called mating. The most fit
individuals reproduce more than the least fit ones. The offspring
(children) form a new population (the second generation). This
reproduction and evaluation process repeats itself until
terminated. The fittest current members of the population survive
until the end.

In our experiments, each individual corresponds to a pulse shape,
which is encoded as a string of floating point numbers specifying
the phase and amplitude at the various frequency components of
the laser pulse. The algorithm typically controls the phases for
$20$ to $60$ colors, linearly interpolating the phases for colors
between the specified frequencies. Unlike the phases, which vary
continuously, the amplitudes are set at discrete levels (between
$3$ and $40$). The actual number of amplitude genes was usually
less than the number of phase genes (between $10$ and $20$). For
some of the experiments, the learning algorithm is only allowed to
vary the phases of the frequency components. This fixes the
energy in each pulse shape, and the algorithm simply determines
how to best distribute this energy in time.

Our population normally consists of $60$ individual pulse shapes.
The first generation is composed of random individuals. Each
pulse shape in the population interacts with the physical system
under investigation and is evaluated for fitness. Our learning
algorithm implements a process known as roulette wheel selection,
in which an individual's chances of reproduction are proportional
to its fitness, so that more fit individuals reproduce more
often. In addition, a given number of the most fit parents are
passed on to the next generation. This technique, known as
elitism, ensures that good genetic material is not lost if by
chance one of the best individuals is not chosen for reproduction
or does not reproduce fruitfully.

\subsection{Operators}

The protocol for the production of new children is carried out by
mathematically defined {\it operators} that act on the gene
strings of the pulse shapes. A traditional GA uses a mixture of
{\it crossover} and {\it mutation} operators, both of which are
described below.  We have incorporated new operators to include
many different methods for searching the available phase space.
The algorithm then combines the entire set of mating operators
into a pool, and the various operators are allowed to compete
against each other for the chance to produce new pulse shapes.
This learning strategy determines which combination of methods is
best for solving the particular problem.

We can categorize our operators as traditional GA operators
(crossovers and mutations), or nontraditional operators.
Traditional operators generate new pulse shapes from old ones on
a strictly statistical basis. Traditional operators that we use
include {\it two-point crossover}, {\it average crossover}, {\it
mutation}, and {\it creep}. Nontraditional operators generally
search the phase space by modifying the pulse shapes in a way
that is guided by the physics of the system. Examples of these
operators that we use include {\it smoothing}, {\it time-domain
crossover}, and {\it polynomial phase mutation}. A well-chosen
set of operators can greatly enhance the performance of the
learning algorithm. Our learning algorithm is general and not
limited to the operators listed above.  As has been discussed
elsewhere, it is possible to add other operators to the pool, even
including entirely new search algorithms, such as simulated
annealing \cite{adler}. If simulated annealing happens to be the
method best suited to the particular problem, the learning
algorithm will discover this, and allow simulated annealing to
control the reproduction process.

Appendix A gives the mathematical form of each of the operators
that we use. A basic operator employed in most traditional GA's
is known as {\it n}-point crossover. This operator selects a
portion of the gene string from each of two (or more) parent
pulse shapes, and then exchanges this section of the gene string
between the two parents. The resulting pair of gene strings are
the two new children pulse shapes. We use a two-point crossover,
which snips the gene string at two random locations and exchanges
the genetic information of the two parents between these two
locations in order to produce two children. Average crossover
also selects two parent gene strings, but rather than exchange
genetic information between them, it averages the values of all
the genes from the two parents to produce a single child. One
important difference between average crossover and two-point
crossover is that average crossover can introduce new gene values
that were not present in either parent, while two-point crossover
simply exchanges values between the two parents.

Another traditional operator that we employ is mutation. This
creates a single child from a single parent by randomly
reassigning the values of a group of randomly selected genes.
Creep, an operator similar to mutation, reassigns values for
selected genes incrementally, adjusting the previous values by a
small but random amount. Like average crossover, both creep and
mutation can introduce new gene values into the population.

Our nontraditional operators only act on the phase genes of the
pulses, leaving the amplitudes fixed.  One of the nontraditional
operators we use is smoothing. The smoothing operator creates a
new pulse from a single parent by performing a three-point
windowed average over the phase values in the gene string. This
operator works very well for problems that require smooth phase
profiles across the bandwidth, and it also aids in the
interpretation of the results because it produces pulses that are
not plagued as much by the entropically driven variations in gene
values that sometimes arise from the GA. The action of smoothing
as seen in the time domain is to shorten the pulse and reduce
structure at long time.

Time-domain crossover is a variation of two-point crossover that
first transforms the gene string into a time domain
representation of the pulse by performing an inverse fast Fourier
transform (IFFT). It then performs a standard two-point crossover
on the time domain pulse representation and transforms the pulse
back to the frequency domain via a FFT. This operator is useful
for problems that involve time domain correlations in the pulse.
High order processes that depend sensitively on $I\left( t\right)
$ are examples of such problems, and we have shown that the  time
domain crossover operator performs well in these situations.

Polynomial-phase mutation produces children by replacing a
portion of a gene string with a polynomial phase function with a
small degree of random variation. The resulting phase profile
resembles a polynomial curve over a section of the spectrum. This
operator and smoothing work well in conjunction to produce pulses
with smooth polynomial phase, which have simpler interpretations
in a time-frequency (i.e., Wigner or Husimi) representation
\cite{paye}.

\subsection{Adaptive Operators}

The adaptive algorithm determines how to best solve a problem by
evaluating ``operator fitnesses''. Like the individual pulse
shapes, each of the operators in the pool is evaluated to form a
basis for operator selection; each operator is chosen to produce
new pulse shapes with a probability proportional to its own
fitness. The operator fitnesses are controlled by compiling an
operator geneology to keep track of the operators responsible for
creating each individual, and assigning a ``credit'' anytime an
operator produces a very fit new pulse shape. Thus, operators
that produce good pulses are given the opportunity to produce more
children. Specifically, credit is assigned when either (1) an
individual pulse shape in the current generation is more fit than
the best pulse shape of the previous generation, or (2) an
individual was an ancestor of a pulse shape that is more fit than
the best pulse shape of the previous generation. Passing credit
back more than one generation is important, since some operators
tend to act in concert. For example, mutation may change genes in
ways that are not beneficial until combined with certain other
genes in the gene string. By combining the changes made by
mutation, two-point crossover may be able to produce fitter
children. Passing the credit back a generation or two insures
that the mutation operator is also rewarded for its contribution.

We begin our algorithm by assigning each operator an initial
normalized fitness, or weighting. The initial weightings are
determined by various methods, including empirically (from both
experiments and simulations) or through prior knowledge of
typical algorithm performances. After the first three
generations, the operator fitnesses are allowed to evolve. At
this point, the weight for each operator has two parts: a base
weight (the value of the operator's fitness during the previous
generation) and an adaptive weight (the operator's fraction of the
total credit assigned to all individuals during the previous
three generations). After each subsequent generation, each
operator's new fitness is a weighted sum of its current base
weight ($85\%$) and its current adaptive weight ($15\%$). Adaptive
weighting allows the operators that produce better children to
increase their operator fitness. This serves two functions. First,
this process speeds up the convergence of the algorithm, since
the operators that are not producing better children, and
therefore not helping the evolution, are prevented from
dominating the reproduction process. Second, the fitnesses of the
different operators can yield insight into the dynamics of the
learning algorithm and also into the physical system itself by
monitoring each operator's fitness as a function of the
generation. Possible control mechanisms can be tested by
introducing new physically motivated operators and evaluating
their success. However, the reproductive process is not required
to ``know'' about the particular goal for the problem. The success
of the algorithm relies on the fact that the adaptive pool of
operators searches vast regions of phase space efficiently,
finding successful individuals without any prior knowledge.

The full power of the learning algorithm is best put to use in
problems where the experimental knobs are all coupled - when the
different degrees of freedom are not independent of each other.
In our experiments, the phases and amplitudes of the individual
colors are coupled by the system Hamiltonian. In a completely
decoupled basis, the problem is reduced to a series of simple
one-dimensional searches for each of the genes. However, one does
not know what this basis is in general. In a coupled basis,
crossover has been demonstrated to be a very valuable operator in
selecting new pulse shapes that out perform their parents.

\section{Experiments}

\subsection{Preliminary Test of Learning Feedback: Second Harmonic Generation in BBO}

Many of the characteristics of adaptive learning are demonstrated
by the simple control experiment of second-harmonic generation
(SHG). Feedback experiments using SHG have previously been carried
out with the goal of targeted pulse compression or shaping
\cite{silberberg-josab,gerber-apb}. Our primary goal was to
investigate the dynamics of our algorithm and the learning process
in a well-characterized and well-studied system. Frequency
doubling in a noncentrosymmetric crystal with a large $\chi
^{\left( 2\right) }$ provided us with this opportunity. Even
though the interaction of the light field with this system can be
described classically, this experiment illustrates features that
are relevant to all the experiments described in this paper. In
addition we are able to simulate the experimental feedback
signal, which allows us to compare learning algorithms on the
model with experiment.

In the low-intensity regime, the interaction of the laser pulses
with the crystal can be described by a nonlinear polarization
that is proportional to the square of the input light field:
\begin{equation}
P_{NL}\left( 2\omega \right) =\chi ^{\left( 2\right) }E\left(
\omega \right) ^{2}.
\end{equation}
This nonlinear polarization acts as a source or driving term in
the wave equation for a field at $2\omega $. The experimental
feedback signal is the integrated second-harmonic intensity:
\begin{equation}
Signal=\int dtE_{2\omega }^{2}\left( t\right)\propto\int
dtE_{\omega }^{4}\left( t\right).
\end{equation}

If the input field strength is not too large, this description of
the interaction gives accurate predictions for the second-harmonic
generation without including other nonlinear effects. For all of
the frequency doubling experiments and simulations, the amplitude
of each frequency component in the pulse is kept fixed. Only the
phases may vary. The laser pulse energy is therefore constant,
and the algorithm determines how to distribute this energy in
time.

The optimal pulse shapes are best viewed as Husimi distributions.
The Husimi distribution, $Q(t,\omega)$ is calculated from the
measured field $E(\omega)$ in the frequency domain:
\begin{equation}
Q(t,\nu)=\int \int dt' d\nu' S(t',\nu')
e^{-(\nu-\nu')^{2}-(t-t')^{2}},
\end{equation}
\begin{equation}
 S(t,\nu)=\int E(\nu+\nu')
E^{\ast}(\nu-\nu')e^{2i\nu' t} d \nu'.
\end{equation}
$S(t,\nu)$ is the Wigner function whose marginals represent the
power spectrum $P(\nu)$ and the temporal intensity $I(t)$ of the
laser pulse:
\begin{equation}
\int d\nu' S(t,\nu')=I(t),
\end{equation}
\begin{equation}
\int dt' S(t',\nu)=P(\nu).
\end{equation}

Husimi distributions are generated using the values of the phase
and amplitude that the pulse shaper programs onto each individual
pulse shape. We use spectral interferometry on a limited number
of pulse shapes to verify the correspondence between the phase
and amplitude profile of the pulse and its representation on our
pulse shaper. Since each convergence run of the algorithm has the
potential to provide a vast amount of information regarding the
problem under study, we monitor not only the optimal pulse shape
solution, but also compare this solution to other competing
solutions from individual generations throughout the run.

\begin{figure}[tb]
\centering
\includegraphics [width=8.6cm, height=8.6cm]{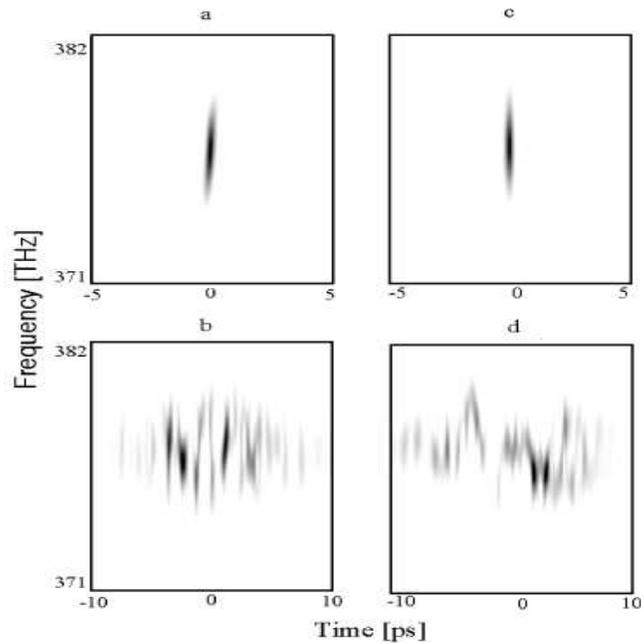}
\caption{Husimi distributions of the optimal pulse shapes for SHG
in BBO.  Panels a and b are experimental results for maximization
and minimization of SHG, respectively. Panels c and d are the
optimal pulse shapes generated using a simulation modeling the SHG
process. The value of the Husimi function is indicated by the
darkness of the shading.} \label{BBO-results}
\end{figure}

Figure \ref{BBO-results} shows the Husimi distributions for
pulses optimized to either maximize or minimize frequency
doubling in BBO. The results of both simulations and experiments
are shown for comparison. Comparison of panels a and c reveals
that experiment and simulation arrive at the same result for
maximizing SHG, except for a small amount of quadratic dispersion
(chirp) evident in the experimental result that is within our
measurement resolution of the laser pulse.

When the algorithm minimizes the SHG, solutions from both the
simulation and experiment contain structure in $I\left( t\right)
$. We initially expected the solutions for the spectral phase
$\phi \left( \omega \right) $ would contain only the lowest
nontrivial order ($ \phi \left( \omega \right) =k\omega ^{2}$).
Given the AOM's constraint of a maximum allowed phase change
between adjacent frequencies in the light pulse, quadratic phase
is the most efficient single-order phase variation, since it
allows for the greatest amount of total phase variation across the
spectrum. However, the algorithm finds that a single order of
phase is not the most efficient way to minimize $I\left( t\right)$
\cite{malinovsky}. We evaluated pulses that were simply chirped in
time by programming them with the maximum amount of quadratic-only
phase allowed by the resolution of our pulse shaper, and found
that they did not perform as well as the solutions found by the
learning algorithm (see Fig. \ref{BBO-results}).

\begin{figure}[tb]
\centering
\includegraphics [width=8.6cm, height=8.6cm, angle=90]{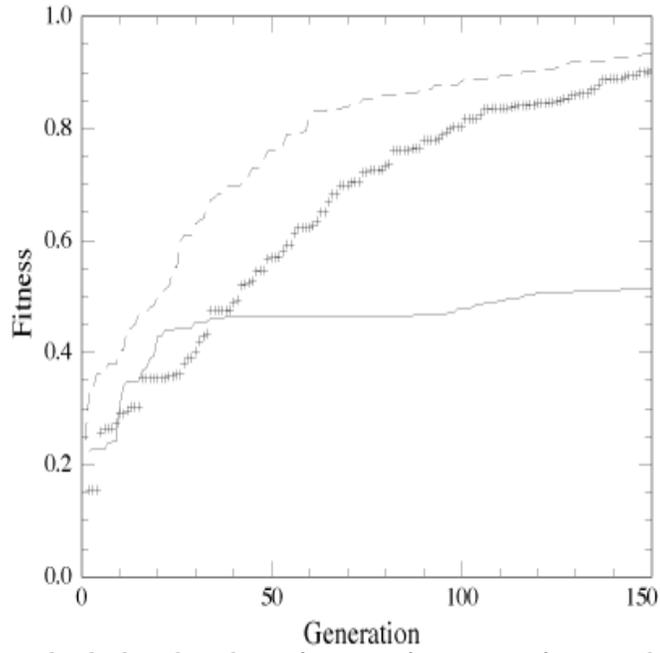}
\caption{The fitness of the best individual is plotted as a
function of generation for a simulation of SHG maximization. The
upper two curves, the dashed and plus lines, compare two runs of
the learning algorithm using all the same operators (including
{\it smoothing}), but starting with different random initial
populations.  The solid line is a third run without the {\it
smoothing} operator.} \label{smoothing-fitness}
\end{figure}

Simulations provide rapid testing of the performance of many
possible operators. For instance, Fig. \ref{smoothing-fitness}
shows the best fitness as a function of generation for three
different runs of the SHG simulation, both with and without the
smoothing operator. The addition of the smoothing operator allows
the algorithm to achieve higher fitness more rapidly and converge
sooner. In the absence of noise, the best fitness increases
monotonically as a function of generation, as guaranteed by
elitism.

\begin{figure}[tb]
\centering
\includegraphics [width=8.6cm, height=8.6cm]{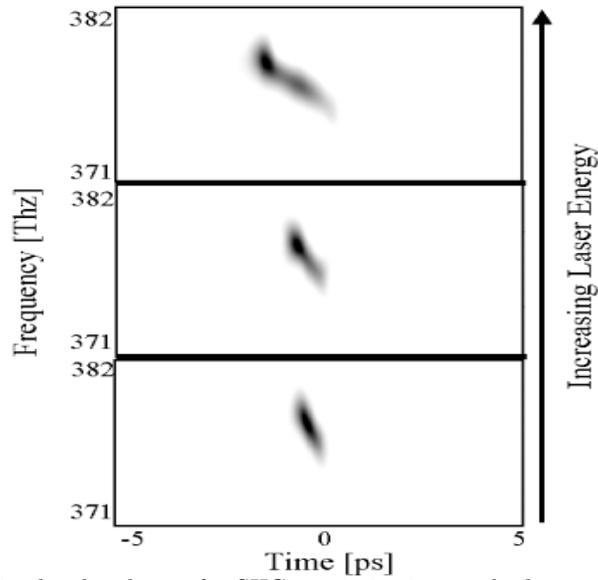}
\caption{Husimi plots of the optimal pulse shapes for SHG
maximization as the laser energy is increased.  At higher input
energies (upper panels), the incident pulse contains higher orders
of dispersion that broaden the pulse in time.} \label{BBO-scan}
\end{figure}

When the intensity of the light increases, our simple model is no
longer adequate to describe the doubling process. Figure
\ref{BBO-scan} shows the changing optimal pulse shape for
maximization of SHG as the energy of the input pulse is
increased. The optimal pulse in the high-energy solutions acquires
large third- and fourth-order dispersion. The shaped pulse
spectrum at high intensities shows significant self-phase
modulation (SPM), whereas in the low intensity limit there is no
evidence of SPM. Simulations are also consistent with SPM in the
crystal, which contributes to variations in the solutions as the
intensity is increased \cite{french-paper}. At higher pulse
energies, SPM is no longer negligible. Evidently, SPM distorts
the phase matching for the SHG process, so that a transform
limited input pulse is no longer optimal for maximum SHG.

\begin{figure}[tb]
\centering
\includegraphics [width=8.6cm, height=8.6cm]{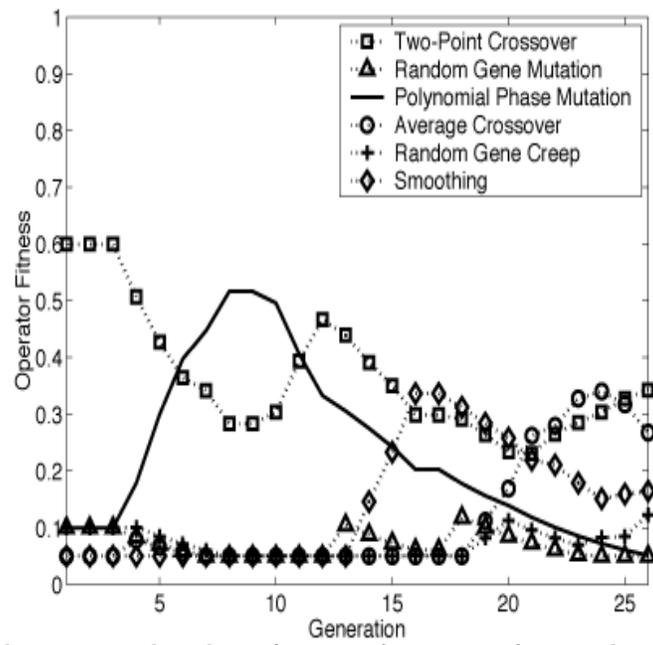}
\caption{The fitness for multiple operators plotted as a function
of generation for a single run of the learning algorithm while
maximizing SHG.} \label{operator-fitness}
\end{figure}

The interplay of multiple operators during the algorithm can yield
further insight into the learning process. Figure
\ref{operator-fitness} shows the operator fitness as a function
of generation for several of our common operators during the
frequency doubling experiment. The operators are initialized in a
traditional GA configuration, with two-point crossover dominant.
For the first few generations, the operators create children in
proportion to their initially assigned fitnesses, but after the
third generation their fitnesses are allowed to freely change in
accordance with the procedure described earlier. As Fig.
\ref{operator-fitness} demonstrates, the algorithm finds that
two-point crossover and simple mutation are not always the best
operators, and at different points during the evolution,
different combinations of operators become optimal for producing
the best children. Polynomial phase mutation, smoothing, and
average crossover each produce very fit children at different
stages of the run.

This example shows that operators cannot be evaluated in
isolation, because they affect each other. For example, smoothing
is more important when other operators tend to introduce
unnecessary phase variations across the pulse. Also, the
performance of the operators cannot be evaluated instantaneously,
but must be evaluated over the course of several generations. An
operator that is not performing well at one point during the run
may become more useful later on. Since some operators that perform
poorly at the beginning of the algorithm often perform very well
toward the end (e.g., average crossover), their fitness is not
allowed to fall below some minimum value ($5\%$). This lower
bound ensures that every operator always has some chance of being
used during reproduction. Finally, the performance of each
operator depends on the problem. Performance of a given operator
can help determine whether its action on an individual is
physically relevant.

The fitness function assigns each individual pulse shape a
single-valued number, reflecting that individual's ability to
achieve the goal. Since an individual's fitness is used in parent
selection (see Appendix B for details), its determination is an
important step in the performance of the algorithm. For the
frequency doubling experiment, the fitness assigned to each
individual was simply the integrated blue light intensity as
measured by a photodiode in a regime where the response of the
diode was linear. The frequency doubling experiment, therefore,
provides a clear testing ground for the learning algorithm. In
other experiments, it is not always so clear how to assign a
fitness to each individual, given the nature of the measurement to
evaluate the success of each pulse.

\subsection{Controlling Dissociative Ionization in Diatomic Sodium}

We next explore nonlinear control mechanisms for ionization of
diatomic sodium. This experiment provides further opportunities
to understand the learning algorithm. However, unlike
second-harmonic generation, which can be described classically,
this is strictly a quantum system. Here, the adaptive algorithm
and pulse shaper must control higher-order nonlinearities. The
multiphoton ionization of this system was previously studied
using low-energy laser pulses with a single photon resonance
enhancement \cite{baumert}. Our experiments use lower-energy laser
photons that are below this resonance.

The shaped pulses were focused into a molecular beam of sodium,
causing the molecules to undergo multiphoton ionization. A
time-of-flight mass spectrometer allowed identification of
molecules that dissociatively ionized and those that did not. The
fitness function was a normalized ratio of the ion yield for the
two channels.  The learning algorithm then worked to optimize
either dissociative or nondissociative ionization. As in the
doubling experiment, we restricted the algorithm to control only
the phases of the colors, with the amplitudes fixed. The algorithm
was able to find pulse shapes that could maximize either channel.
In the case of maximizing the nondissociative channel, the
optimal pulse yielded $88\%$ nondissociative ionization. When we
optimized the other channel, we found $73\%$ dissociative
ionization.

\begin{figure}[tb]
\centering
\includegraphics [width=8.6cm, height=17.2cm]{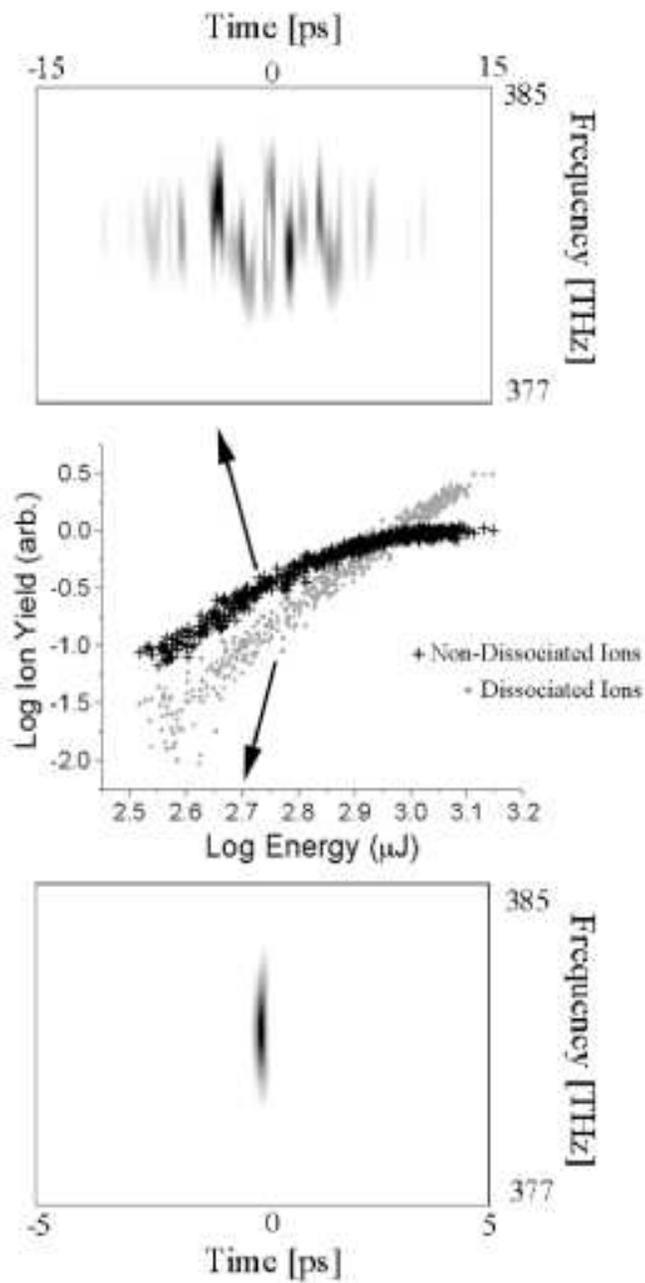}
\caption{Husimi distributions of the optimal pulse shape for
controlling sodium ionization.  The upper plot is for maximizing
the nondissociative yield, while the lower plot is for the
dissociative yield. In between is a log-log plot of the ion yield
into each of the two channels as a function of the energy of an
unshaped laser pulse.} \label{sodium-results}
\end{figure}

Figure \ref{sodium-results} shows the Husimi distributions for
pulses optimized to either nondissociatively or dissociatively
ionize the sodium molecules. In addition, the ion yield as a
function of the laser energy for an unshaped laser pulse is
plotted for both of the channels. The solutions resemble the
optimal pulse shapes for the second-harmonic generation experiment
(see Fig. \ref{BBO-results}). For the dissociative channel, the
pulse shape is similar to the short-pulse, high-intensity
solution of SHG maximization, while for the nondissociative
channel, the pulse shape is similar to the long-pulse,
low-intensity solution of SHG minimization. These results are
consistent with the fact that the two channels in log-log plot
have different slopes, and that the branching ratio is a function
of intensity.

\begin{figure}[tb]
\centering
\includegraphics [width=8.6cm, height=8.6cm]{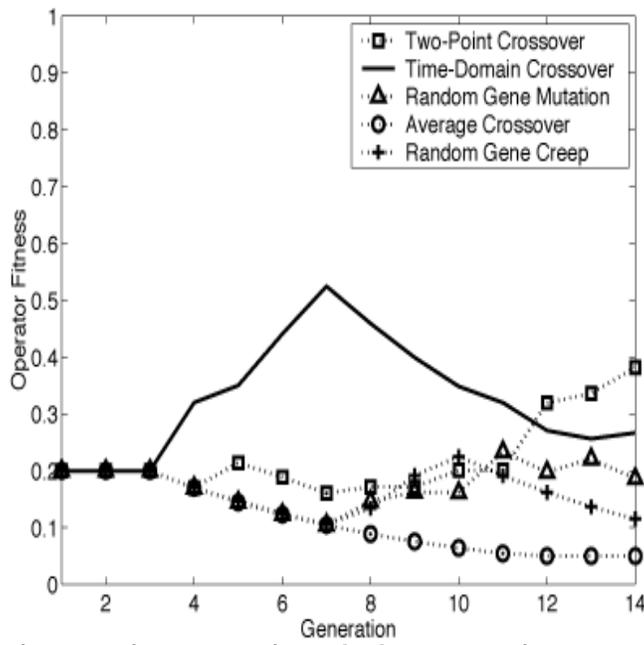}
\caption{Operator fitness as a function of generation for
multiple operators for optimizing the nondissociative ionization
channel of sodium.} \label{time-op-fitness}
\end{figure}

The sodium experiment demonstrates the importance of the choice of
basis for encoding pulse shapes. The learning algorithm
incorporates the time-domain crossover operator, which allows it
to choose the basis best suited to the problem. Figure
\ref{time-op-fitness} shows the operator fitnesses as a function
of generation when optimizing the nondissociative channel. There
are five operators - four in the usual frequency basis, and the
time-domain crossover operator in the time basis. As Fig.
\ref{time-op-fitness} shows, the time-domain crossover operator
increases its fitness at the expense of the operators working in
the frequency basis. Correlations in the time domain are an
important control parameter for this problem, as expected.

Another possible control mechanism involves the parity of the
phase profile of the pulse: Is the phase as a function of
frequency an even function, an odd function, or neither about the
central frequency? Parity control in nonlinear atomic absorption
was previously demonstrated by Meshulach and Silberberg
\cite{silberberg-nature}. The amount and sign of chirp on the
pulse is another possible way to control the population dynamics
\cite{wilson}. We investigated both possibilities using our pulse
shaper and feedback, but saw no conclusive dependence on the
dissociation fraction with either method. The adaptive pool of
operators performs a more effective search of the phase space.

\subsection{Controlling Molecular Liquids}

Molecular liquids pose a greater challenge for learning control,
because of rapid relaxation and inhomogeneous broadening. Our
initial goal was to control molecular vibrations in liquids
through impulsive stimulated Raman scattering. Impulsive
scattering occurs when the laser pulse is shorter than the
vibrational period of the molecule. In the frequency domain, this
means that the bandwidth of the laser is broad compared to the
vibrational energy spacing of the mode in question.  The result
is that a stimulated Stokes wave can be seeded with light that is
already present in the laser and does not have to build up from
noise, thereby making the process much more efficient than the
nonimpulsive case. Shaped laser pulses should allow selective
control over excitation of Raman modes since the pulse shaper can
modulate the spectrum of $E^{2}(\omega)$, the driving term in
Raman excitation \cite{news+views}. In the time domain, this
corresponds to resonantly driving some modes but not others.

\subsubsection{Controlling SPM in Liquids}

Impulsive scattering in a multimode molecular liquid was studied
in $CCl_{4}$ because it has several low frequency modes with
relatively high cross sections. We discovered, however, that it
also has a large polarizability, and therefore, most of the light
that was scattered near the laser bandwidth in the forward
direction was a result of SPM. We found that we had a significant
degree of control over the spectrum of the forward scattered
light, and so as an initial demonstration of the capabilities of
the apparatus, we studied the nonlinear frequency shift of intense
light propagating in $CCl_{4}$. Early results of this
investigation were previously published \cite{jpc}. Here, we
describe the learning process and analyze the solutions in more
detail.

\begin{figure}[tb]
\centering
\includegraphics [width=8.6cm, height=8.6cm, angle=90]{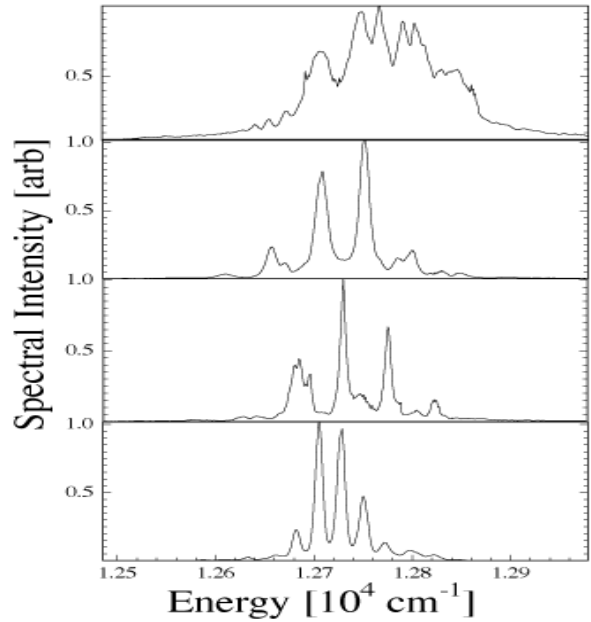}
\caption{Power spectra for pulses propagating through $1 cm$ of
$CCl_{4}$. The top panel shows the power spectrum for an unshaped
laser pulse. The following three panels show spectra for pulses
shaped to control (and enhance) the spectral modulations. (From
Ref. \protect\cite{jpc}, reprinted by permission)}
\label{ccl4-spectra}
\end{figure}

Feedback goals for SPM are based upon small features that are
barely visible in the spectra of the forward scattered radiation
after unshaped pulses illuminate the sample. These modulations
are typical spectral features for pulses that have undergone SPM
\cite{corkum-spm}. The learning algorithm is able to control their
frequency and phase by altering the shape of the driving pulse.
Only phase modulation is used, so that the pulse energy is fixed.
Figure \ref{ccl4-spectra} shows the spectra of four different
pulses after propagating through the $CCl_{4}$ sample. The first
panel shows the spectrum for an unshaped pulse, and the following
three panels show spectra for pulses that are shaped to control
the spectral modulations.

Model calculations can determine whether SPM is responsible for
the observed spectra. The simplest description of SPM, which
doesn't include spatial effects such as self focusing,
characterizes the nonlinear interaction between the laser and the
medium through a intensity-dependent index of refraction:
\begin{equation}
n(t)=n_{0}+n_{2}I(t)
\end{equation}
where $n_{0}$ is the field free index, I is the instantaneous
laser intensity, and $n_{2}$ is an empirically determined
coefficient. The calculated power spectra for laser pulses that
have acquired a phase proportional to their instantaneous
intensity show intensity modulations with the same dependence on
pulse shape that we found in the experiment (see Fig.
\ref{ccl4-spectra}).

\subsubsection{Controlling Vibrations in Multimode Molecular
Liquids}

The learning algorithm can also control the interaction between
the driving laser pulse and the vibrational modes of a multimode
molecule without making use of impulsive scattering. In order to
avoid confusion between Stokes light and light generated by SPM
alone, we chose a molecule with a much larger Stokes shift, since
there is much less light generated through SPM further from the
central laser frequency. Methanol ($CH_{3}OH$) is ideal because
it has two closely spaced modes with large Stokes shifts and
large cross sections. Forward scattered radiation is the feedback
for the algorithm. There is no backwards scattered Stokes
radiation because of the short duration of the shaped pump pulse
($\sim1ps$). The forward-backward symmetry of the scattering is
broken for a short pump pulse because the backward traveling
Stokes wave passes through the pump wave before any appreciable
buildup \cite{shen}.

The time scales for the interaction between the molecules and the
laser pulse are set by the vibrational period of the active modes
and their coherence time. Stimulated scattering with pulses that
are longer than the coherence time reaches a steady state and
exhibits a strong dependence on pulse duration because it is a
stimulated process: The more photons that interact with the
molecules within the coherence time, the more likely the molecule
will be stimulated to absorb a laser photon and emit a Stokes
photon. Scattering with pulses that are shorter than the
coherence time but longer than the vibrational period (transient
Raman scattering) exhibits little dependence on the duration of
the laser pulse \cite{bloembergen-theory,carman}. Scattering with
pulses that are shorter than the vibrational period (impulsive
scattering) results in very efficient stimulated scattering as
discussed above \cite{korn-theory,korn-experiment}.

\begin{figure}[tb]
\centering
\includegraphics [width=8.6cm, height=8.6cm, angle=90]{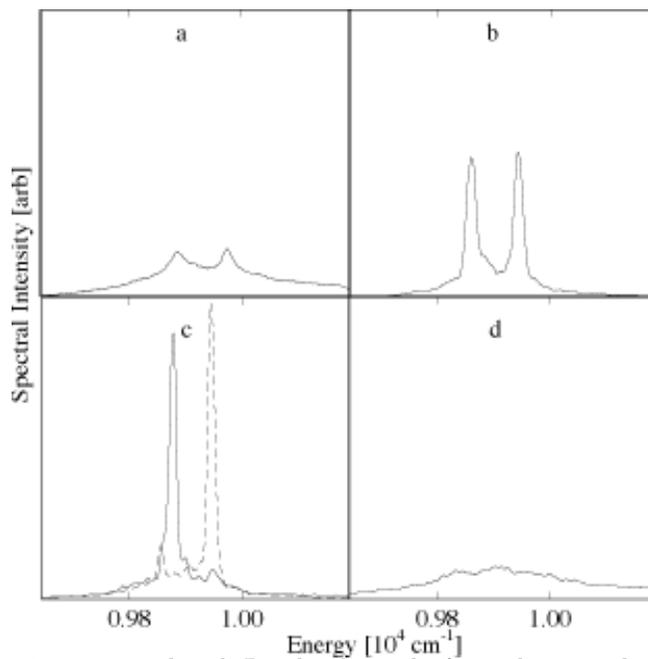}
\caption{Control of Raman scattering in methanol. Panel a shows
the forward scattered spectrum for an incident unshaped laser
pulse. Panel b shows spectrum after the learning algorithm
optimized excitation of both modes while minimizing peak
broadening due to other nonlinear effects. Panel c shows spectra
for optimization of each mode independently. Panel d shows
spectrum for a pulse that minimized Raman scattering from both
modes. (From Ref. \protect\cite{jpc}, reprinted by permission)}
\label{meth-peaks}
\end{figure}

With an unshaped laser pulse focused into $10cm$ of methanol, the
spectrum shown in Fig. \ref{meth-peaks} a is obtained. The two
small peaks in the spectrum correspond to Stokes light for
$\delta \nu =1$ for the symmetric and asymmetric C-H stretch
modes. The lens focal length is $40cm$. The first feedback goal
is to maximize the contrast between the two Stokes peaks and the
background light resulting from SPM. The forward scattered
spectrum in the spectral range of the Stokes radiation is
collected for each laser pulse. The number of Stokes photons at a
particular frequency is a measure of the number of molecules
excited in that particular mode; however, since SPM is present
and the bandwidth of the shaped laser pulses is large, the
forward scattered spectrum also contains some misleading
information. The feedback function must filter this out.
Different fitness functions work best, depending on which
peak(s)are optimized. A typical fitness function is:
\begin{equation}
\sum_{\omega _{r}<\omega _{i}<\omega _{b}}\frac{N C\left( \omega
_{i}\right) }{\omega _{b}-\omega _{r}}-\sum_{\omega _{i}>\omega
_{b},\omega _{i}<\omega _{r}}\frac{C\left( \omega _{i}\right)
}{\Delta \omega -\left( \omega _{b}-\omega _{r}\right) }.
\end{equation}
Here, $C\left( \omega _{i}\right)$ is the number of spectrometer
counts at $\omega_{i}$, $\Delta \omega$ is the bandwidth of the
spectrometer, $\omega _{r}$ is the low-frequency limit for the
desired peak, $\omega _{b}$ is the high-frequency limit for the
desired peak and N is an empirically determined integer. We set
the values of $\omega_{r}$ and $\omega_{b}$ by narrowing the
bandwidth of shaped laser pulse and measuring the width of the
Stokes peaks in the forward scattered spectrum.  We find that
$N=2$ and $N=3$ work well. A pulse solution that optimizes the
contrast between the two Stokes peaks and the background is shown
in Fig. \ref{meth-peaks} b.

The next goal is to generate spectra with each peak separately.
These spectra, shown in Fig. \ref{meth-peaks} c, correspond to
exciting symmetric or antisymmetric modes alone. The Stokes shift
for these modes is large compared to the bandwidth of the driving
laser pulse. This is equivalent to saying that the Raman
excitation is nonimpulsive, and therefore, one cannot seed the
Stokes radiation directly with the laser light. Our final
feedback goal is to eliminate all forward scattered light at
either of the two Stokes frequencies, with the resulting spectrum
shown in Fig. \ref{meth-peaks} d.

\subsubsection{Using the Adaptive
Algorithm to Investigate Possible Control Mechanisms for SRS in
Methanol}

The Stokes shift for the C-H stretch is almost $3000 cm^{-1}$and
the laser bandwidth is roughly $100 cm^{-1}$, so the scattering is
definitely nonimpulsive. Another possible mechanism that could
account for our ability to selectively excite the symmetric or
asymmetric stretch mode of methanol is a coupling between the
electronic polarizability of the atoms and the vibrational modes.
The results with $CCl_{4}$ have shown that the light generated
from SPM of the pump beam is very sensitive to the input pulse
shape. A large contribution to SPM, particularly for femtosecond
pulses, is the atomic polarizability \cite{ccl4}. Perhaps the
atomic polarizability generates SPM to seed one of the two Raman
modes but not the other.

\begin{figure}[tb]
\centering
\includegraphics [width=8.6cm, height=8.6cm, angle=90]{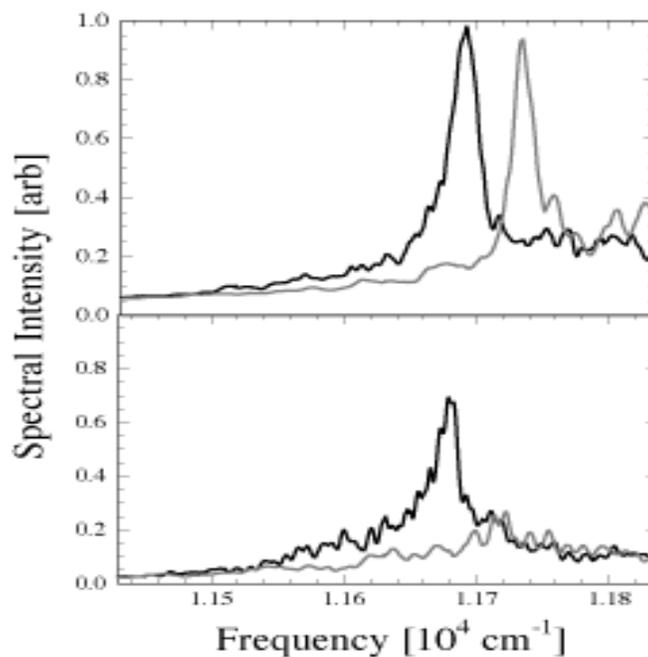}
\caption {Stimulated Raman scattering in $C_{6}H_{6}$ and
$C_{6}D_{6}$. Top panel shows spectra for $C_{6}H_{6}$ (left
curve) and $C_{6}D_{6}$ (right curve) separately after
optimization of the pulse shape to excite the breathing mode of
each molecule. Bottom panel shows the results after using the
learning algorithm to excite each molecule separately in a
$50/50$ mixture of the two.} \label{benzene-eps}
\end{figure}

This hypothesis was tested by replacing the methanol with a
mixture of benzene ($C_{6}H_{6}$) and deuterated benzene
($C_{6}D_{6}$). Similar control experiments were attempted, except
now the two modes of vibration were in two different molecules.
The ring breathing mode of benzene $ \left( \nu
=992cm^{-1}\right) $ has a large Raman cross section making
excitation easy.  Deuterated benzene has a frequency of $\nu
_{D}=945cm^{-1}$ which is shifted by $47cm^{-1}$, similar to the
mode splitting in methanol. Initially, an unshaped laser pulse was
focused into the experimental cell with pure $ C_{6}H_{6}$, and
we measured no forward scattered Stokes light. We then used the
learning algorithm to find a shaped pulse that generates the
forward scattered Stokes radiation shown in Fig. \ref{benzene-eps}
(top panel, left curve). The deuterated benzene ($C_{6}D_{6}$)
generates the spectrum shown in Fig. \ref{benzene-eps} (top panel,
right curve), demonstrating the expected shift of the mode
frequency. Finally, a $50/50$ mixture of $C_{6}H_{6}$ and
$C_{6}D_{6}$ is placed in the cell and we ask the algorithm to
selectively drive each of the two modes. Figure
\ref{benzene-eps}(bottom panel) shows the learning algorithm can
select the $C_{6}H_{6}$ mode but not the $C_{6}D_{6}$ mode. This
demonstration of a lack of control is consistent with the idea
that the mode selection is an intramolecular effect that relies
on coupling between the two modes inside each molecule, rather
than seeding of one of the modes with light from SPM.

Another experiment supports this conclusion: Experiments conducted
in $CO_{2}$ gas show that light generated by SPM more than a few
hundred $cm^{-1}$ away from the laser frequency is extremely noisy
and not reproducible \cite{thesis}. The reproducibility of SPM
spectra increases nearer the frequency of the driving laser.
Figure \ref{ccl4-spectra} shows that SPM produces stable spectra
very near the laser frequency. However, far from the laser at the
frequency of the Stokes light from the C-H stretch mode, the
light produced by SPM would be too noisy to reproducibly seed one
of the two Raman modes but not the other.

\begin{figure}[tb]
\centering
\includegraphics [width=8.6cm, height=8.6cm, angle=0]{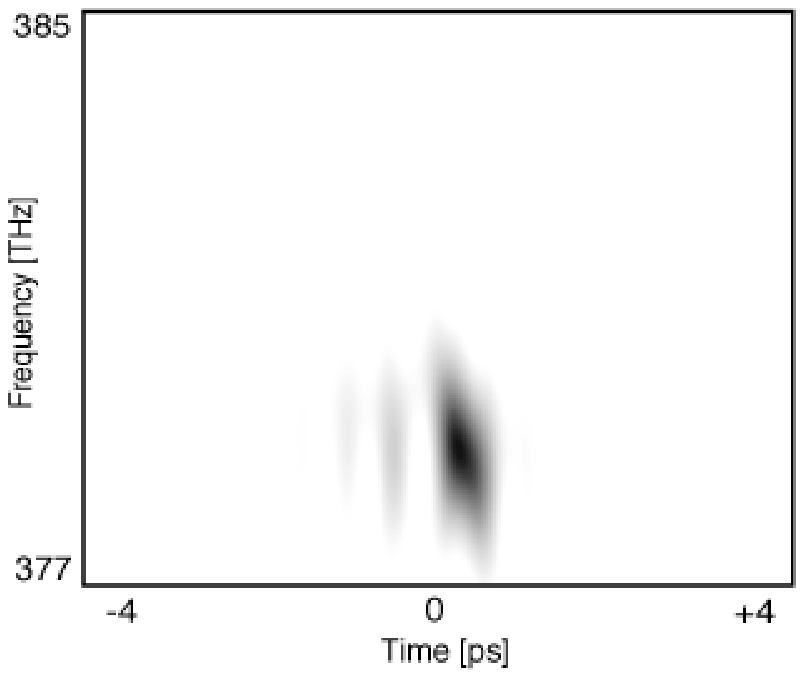}
\caption{Husimi plot of the optimal pulse shape to excite the
asymmetric stretch in methanol.} \label{meth1-husimi}
\end{figure}

\begin{figure}[bt]
\centering
\includegraphics [width=8.6cm, height=8.6cm, angle=0]{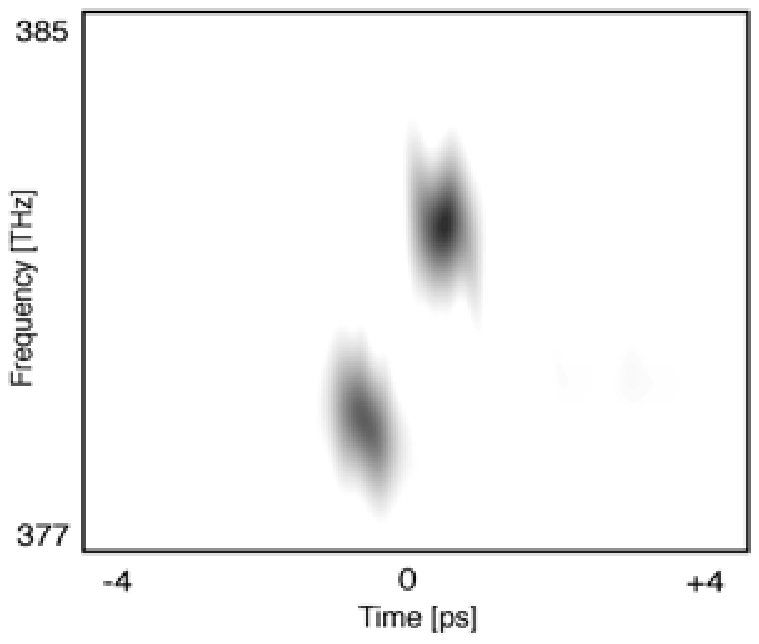}
\caption{Husimi plot of the optimal pulse shape to excite the
symmetric stretch in methanol.} \label{meth2-husimi}
\end{figure}

Another possible control mechanism is suggested by analyzing the
optimal pulse shape solutions for exciting each of the two modes.
Figure \ref{meth1-husimi} shows the Husimi distribution for
pulses that were optimized for excitation of the asymmetric
stretch mode in methanol, while Fig. \ref{meth2-husimi} shows the
Husimi distribution for pulses optimized for excitation of the
symmetric stretch mode. The structure of the optimal pulse shape
for the symmetric stretch mode suggests a ``quasi-impulsive''
model, where the frequency separation of the two subpulses is
exactly the beat frequency between the symmetric and asymmetric
modes. The term ``quasi-impulsive'' is used, since although the
laser bandwidth is narrow compared to the Stokes shift of each
mode, it is wide in comparison to the spacing between the modes.
This is equivalent to saying that in the time domain, the laser
pulse is long compared to the vibrational period of the modes,
but short compared to the beat note between them. The beat note
period is $285$ fs, while the time duration of the unshaped laser
pulse is $150$ fs. Therefore, energy could be transferred between
the two modes by engineering temporal structure in the driving
pulse at the coupling frequency between the two modes. Once an
initial vibrational population is established in some combination
of the two modes, the population could be redistributed by the
shaped pulse through an impulsive coupling of the two levels.

General trends in an entire population can provide valuable
information, since looking at a single individual, even if it is
the very best pulse shape of the group, does not give information
about which features are necessary and which are merely
sufficient. One cannot tell whether a feature in the pulse shape
plays some physical role in the process under investigation, or
whether the algorithm simply did not remove the feature since its
presence did not degrade the fitness. We test this by repeating a
convergence sequence of the learning algorithm with different
random initial populations. If the best solution has similar
structure multiple times, it is likely that the structure is
physically necessary.

\begin{figure}[tb]
\centering
\includegraphics [width=8.6cm, height=8.6cm]{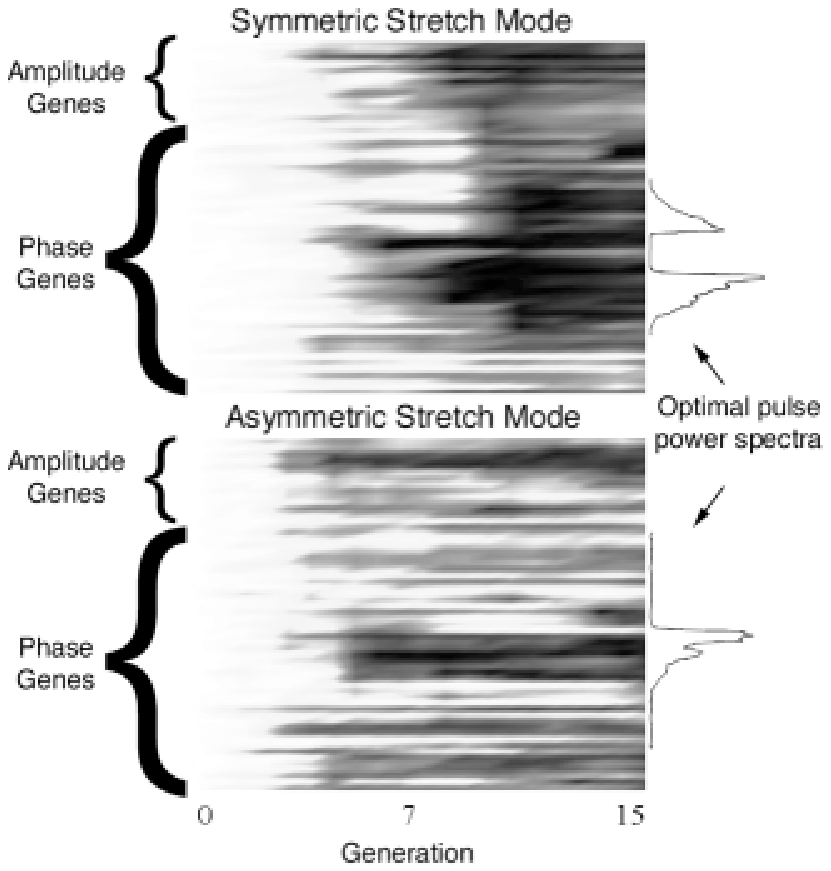}
\caption {Genetic variation as a function of generation. Data is
shown for two separate runs of the learning algorithm.  Top panel
shows the genetic variation for optimization of the symmetric
stretch mode, and the bottom panel for optimization of the
asymmetric stretch mode. The vertical axis represents the gene
number. For reference, the power spectra are overlayed to the
right of the phase genes. Light shading represents a large degree
of variation among the individuals, while dark shading represents
low variation.} \label{gene-spread}
\end{figure}

Statistical variations among gene values of individuals in a
population also reveals which genes are important for a given
problem and which are not. Figure \ref{gene-spread} shows the
genetic variation as a function of generation for a run of the
algorithm during the methanol experiment. Light shading
represents a large degree of variation among individuals for the
value of a given gene, while dark shading represents low
variation.  The variation is the normalized sum of the absolute
values of the differences between all of the genes' values in a
given location on the gene string:
\begin{equation}
\sum_{i,j>i}^{N}|g_{i}-g_{j}|,
\end{equation}
where $g_{i,j}$ is the value of gene $g$ for the $i^{th},j^{th}$
individual. Since the genes are randomly initialized, all the
genes begin with a light shading. As the algorithm converges, all
gene values become more similar through mating. The plots verify
the intuitive idea that near convergence, frequency components
whose amplitudes are large have smaller variation in their
programmed phase values than frequency components whose
amplitudes are zero.  The power spectra shown at right match with
the darker regions (smaller variation) in the plots.

\section{Conclusions}

We have demonstrated control over a variety of systems using an
adaptive learning algorithm.  The algorithm uses a variety of
searching methods and adapts itself in order to arrive at an
optimal solution. This learning technique is general and can be
applied to other systems since the interaction does not require
specific resonances and no prior knowledge of the system
Hamiltonian is required. The adaptive approach provides
information about the physical system through examination of the
solutions, the operator dynamics, and the choice of basis. Future
goals include controlling bond excitation to drive reactions in
bimolecular solutions.

\section*{Acknowledgements}

The authors would like to acknowledge very helpful and
informative discussions with Marcus Motzkus, Ronnie Kosloff,
Misha Ivanov, and Robert Levis, as well as Daniel Morris for
technical assistance. This work was supported by the National
Science Foundation under Grant No. $9987916$.

\appendix
\section{Definitions of Operators}


\[
\text{r}_{i}=rand(), P = parent, C = child
\]

\begin{eqnarray*}
&&\text{Two Point Crossover} \\
P1 &=&\left[ \Phi _{p1}\left( \omega _{i}\right) ,A_{p1}\left(
\omega _{i}\right) \right] , P2=\left[ \Phi _{p2}\left( \omega
_{i}\right),A_{p2}\left( \omega _{i}\right) \right] \\
C1 &=&\left\{
\begin{array}{c}
\left[ \Phi _{p2}\left( \omega _{i}\right) ,A_{p2}\left( \omega
_{i}\right)
\right] ,j<i<k \\
\left[ \Phi _{p1}\left( \omega _{i}\right) ,A_{p1}\left( \omega
_{i}\right) \right] ,\text{all other i}
\end{array}
\right\}, C2=\left\{
\begin{array}{c}
\left[ \Phi _{p1}\left( \omega _{i}\right) ,A_{p1}\left( \omega
_{i}\right)
\right] ,j<i<k \\
\left[ \Phi _{p2}\left( \omega _{i}\right) ,A_{p2}\left( \omega
_{i}\right) \right] ,\text{all other i}
\end{array}
\right\}
\end{eqnarray*}
\begin{eqnarray*}
&&\text{Average Crossover } \\
P1 &=&\left[ \Phi _{p1}\left( \omega _{i}\right) ,A_{p1}\left(
\omega _{i}\right) \right] ,P2=\left[ \Phi _{p2}\left( \omega
_{i}\right)
,A_{p2}\left( \omega _{i}\right) \right] \\
C &=&\left\{
\begin{array}{l}
\left[ \frac{\Phi _{p1}\left( \omega _{i}\right) +\Phi
_{p2}\left( \omega _{i}\right) }{2},\frac{A_{p1}\left( \omega
_{i}\right) +A_{p2}\left( \omega
_{i}\right) }{2}\right] ,j<i<k \\
\left[ \Phi _{p}\left( \omega _{i}\right) ,A_{p}\left( \omega
_{i}\right) \right] ,\text{all other i}
\end{array}
\right\}
\end{eqnarray*}

\begin{eqnarray*}
&&\text{Mutation} \\
P &=&\left[ \Phi _{p}\left( \omega _{i}\right) ,A_{p}\left( \omega
_{i}\right) \right] \\
\text{ }C &=&\left\{
\begin{array}{c}
\left[ 2\pi *r_{i},r_{i}\right] ,\text{i}>\text{k} \\
\left[ \Phi _{p}\left( \omega _{i}\right) ,A_{p}\left( \omega
_{i}\right) \right] ,\text{all other i}
\end{array}
\right\}
\end{eqnarray*}
\begin{eqnarray*}
&&\text{Polynomial Phase Mutation} \\
P &=&\left[ \Phi _{p}\left( \omega _{i}\right) ,A_{p}\left( \omega
_{i}\right) \right] \\
\text{ }C &=&\left\{
\begin{array}{l}
\left[ \omega _{i}^{n}+r_{i},A_{p}\left( \omega _{i}\right) \right] ,j<i%
\text{ or }i<k \\
\left[ \Phi _{p}\left( \omega _{i}\right) ,A_{p}\left( \omega
_{i}\right) \right] ,\text{all other i}
\end{array}
\right\}
\text{ where }n=(int)6*r_{i}
\end{eqnarray*}

\begin{eqnarray*}
&&\text{Creep} \\
P &=&\left[ \Phi _{p}\left( \omega _{i}\right) ,A_{p}\left( \omega
_{i}\right) \right] \\
C &=&\left\{
\begin{array}{l}
\left[ \Phi _{p}\left( \omega _{i}\right)
+0.25*rand(),A_{p}\left( \omega
_{i}\right) +A_{level}*(\pm 1,0)\right] ,\text{ }r_{i}<0.10\text{ } \\
\left[ \Phi _{p}\left( \omega _{i}\right) ,A_{p}\left( \omega
_{i}\right) \right] ,r_{i}\geq 0.10
\end{array}
\right\}
\end{eqnarray*}

\begin{eqnarray*}
&&\text{Smooth} \\
P &=&\left[ \Phi _{p}\left( \omega _{i}\right) ,A_{p}\left( \omega
_{i}\right) \right] \\
C &=&\left[ \frac{\Phi _{p}\left( \omega _{i}\right) +\Phi
_{p}\left( \omega _{i+1}\right) +\Phi _{p}\left( \omega
_{i-1}\right) }{3},A_{p}\left( \omega _{i}\right) \right]
\end{eqnarray*}

\begin{eqnarray*}
&&\text{Time Domain Crossover} \\
P1 &=&\left[ \Phi _{p1}\left( \omega _{i}\right) ,A_{p1}\left(
\omega _{i}\right) \right] ,P2=\left[ \Phi _{p2}\left( \omega
_{i}\right)
,A_{p2}\left( \omega _{i}\right) \right]  \\
P1^{^{\prime }} &=&IFFT\left[ \Phi _{p1}\left( \omega _{i}\right)
,A_{p1}\left( \omega _{i}\right) \right] =E_{1}\left( t_{i}\right)
,
P2^{^{\prime }}=IFFT\left[ \Phi _{p2}\left( \omega _{i}\right)
,A_{p2}\left( \omega _{i}\right) \right] =E_{2}\left( t_{i}\right)  \\
C1^{\prime } &=&\left\{
\begin{array}{l}
\left[ E_{1}\left( t_{i}\right) \right] ,j<i\text{ or }i<k \\
\left[ E_{2}\left( t_{i}\right) \right] ,\text{all other i}
\end{array}
\right\},
C2^{\prime }=\left\{
\begin{array}{l}
\left[ E_{2}\left( t_{i}\right) \right] ,j<i\text{ or }i<k \\
\left[ E_{1}\left( t_{i}\right) \right] ,\text{all other i}
\end{array}
\right\}  \\
C1 &=&IFFT\left( C1^{\prime }\right), C2=IFFT\left( C2^{\prime
}\right)
\text{but original amplitudes are kept since} A\left(
\omega _{i}\right) \leq 1
\end{eqnarray*}

\section{Details of Fitness Function}

Each individual is evaluated, and a single-valued fitness is
returned to the algorithm. Each individual's fitness is
transformed into a scaled fitness used for parent selection during
reproduction. The fitness scaling helps to ensure there is an
adequate degree of ``selection pressure'', which is a measure of
how much successful pulse shapes are rewarded. Maintaining
selection pressure becomes difficult when the range of fitness
values decreases as the algorithm converges. Without fitness
scaling the algorithm can stagnate since more fit pulse shapes
receive little reward. Specifically, we use a linear scaling
technique:

\[
Scaled Fitness = \frac{Best Fitness - Unscaled Fitness}{Average
Fitness - Best Ftiness} +2 \] Any scaled fitnesses less than zero
are reset to zero. With this scaling method the best pulse shape
is selected as a parent twice as often as the average pulse shape.


%
%

%
%


\begin{thebibliography}{99}

\bibitem{rabitz}  R. Judson and H. Rabitz, Phys. Rev. Lett. {\bf
68}, 1500 (1992).

\bibitem{gordon}  L. Zhu, V. Kleiman, X. Li, S. Lu, K. Trentelman,
and R. J. Gordon, Science {\bf 270}, 77 (1995).

\bibitem{gerber-science}  A. Assion, T. Baumert, M. Bergt, T. Brixner,
B. Kiefer, V. Seyfried, M. Strehle, and G. Gerber, Science {\bf
282}, 919 (1998).

\bibitem{silberberg-nature}  D. Meshulach and Y. Silberberg, Nature
(London) {\bf 396}, 298 (1998).

\bibitem{wilson}  C. J. Bardeen, V. V. Yakovlev, K. R. Wilson, S. D.
Carpenter, P. M. Weber, and W. S. Warren, Chem. Phys. Lett. {\bf
280}, 151 (1997).

\bibitem{zewail}  J. J. Gerdy, M. Dantus, R. M. Bowman, A. H.
Zewail, Chem. Phys. Lett. {\bf 171}, 1 (1990).

\bibitem{heberle}  A. P. Heberle, J. J. Baumberg, and K. Köhler,
Phys. Rev. Lett. {\bf 75}, 2598 (1995).

\bibitem{bonadeo}  N. H. Bonadeo, J. Erland, D. Gammon, D. Park,
D. S. Katzer, and D. G. Steel, Science {\bf 282}, 1473 (1998).

\bibitem{vandriel}  A. Hach\'{e}, Y. Kostoulas, R. Atanasov, J. L. P.
Hughes, J. E. Sipe, and H. M. van Driel, Phys. Rev. Lett. {\bf
78}, 306 (1997).

\bibitem{GA}  {\it Handbook of Genetic Algorithms}, edited by L.
Davis (Van Norstrand Reinhold, New York, 1991).

\bibitem{specint}  D. N. Fittinghoff, J. L. Bowie, J. N. Sweester, R.
T. Jennings, M. A. Krumbugel, K. W. DeLong, R. Trebino, and I. A.
Walmsley, Opt. Lett. {\bf 21}, 884 (1996).

\bibitem{warren}  J. X. Tull, M. A. Dugan, and W. S. Warren,
Adv. Magn. Opt. Reson. {\bf 20}, 1 (1990).

\bibitem{nature}  T. C. Weinacht, J. Ahn, and P. H. Bucksbaum, Nature
(London) {\bf 397}, 233 (1999).

\bibitem{schwefel}  Hans-Paul Schwefel {\it Evolution and Optimum
Seeking} (Wiley, New York, 1995).

\bibitem{motzkus}  T. Hornung, R. Meier, D. Zeidler, K. L. Kompa,
D. Proch, and M. Motzkus, Appl. Phys. B: Lasers Opt. {\bf 71}, 277
(2000).

\bibitem{adler}  D. Adler, in {\it Proceedings of the 1993 IEEE International
Conference on Neural Networks} (SOS Printing, San Diego, 1993).

\bibitem{paye}  J. Paye, IEEE J. Quantum Electron. {\bf 28},
2262 (1992).

\bibitem{silberberg-josab}  D. Meshulach, D. Yelin, and Y.
Silberberg, J. Opt. Soc. Am. B {\bf 15}, 1615 (1998).

\bibitem{gerber-apb}  T. Baumert, T. Brixner, V. Seyfried, M.
Strehle, and G. Gerber, Appl. Phys. B: Lasers Opt. {\bf 65}, 779
(1997).

\bibitem{malinovsky}  V. S. Malinovsky and J. L. Krause, Phys.
Rev. A {\bf 63}, 043415 (2001).

\bibitem{french-paper} E. J. T. Nibbering, M. A. France, B. A. Prade,
G. Grillon, C. Le Blanc, and A. Mysryowicz, Opt. Commun. {\bf
119}, 479 (1995).

\bibitem{baumert}  T. Baumert, B. Bühler, R. Thalweiser, and G.
Gerber, Phys. Rev. Lett. {\bf 64}, 733 (1990).

\bibitem{news+views}  P. H. Bucksbaum, Nature (London) {\bf 396}, 217
(1998).

\bibitem{jpc} T. C. Weinacht, J. L. White, and P. H. Bucksbaum,
J. Phys. Chem. A {\bf 103}, 10166 (1999).

\bibitem{corkum-spm}  P. Corkum and C. Rolland, IEEE J.
Quantum Electron. {\bf 25}, 2634 (1989).

\bibitem{shen}  Y. R. Shen, {\it The Principles of Nonlinear
Optics} (Wiely, New York, 1984).

\bibitem{bloembergen-theory} R. L. Carman, F. Shimizu,
C. S. Wang, and N. Bloembergen, Phys. Rev. A {\bf 2}, 60 (1970).

\bibitem{carman} R. L. Carman, M. E. Mack, F. Shimizu,
and N. Bloembergen, Phys. Rev. Lett. {\bf 23}, 1327 (1969).

\bibitem{korn-theory} G. Korn, O. Duhr, and A. Nazarkin,
Phys. Rev. Lett. {\bf 81}, 1215 (1998).

\bibitem{korn-experiment} A. Nazarkin and G. Korn,
Phys. Rev. A {\bf 58}, R61 (1998).

\bibitem{ccl4} R. W. Hellwarth, A. Owyoung, and N. George,
Phys. Rev. A {\bf 4}, 2342 (1971).

\bibitem{thesis} T. C. Weinacht, Ph.D. thesis, University of
Michigan, 2000.

\end{thebibliography}
\end{document}